\documentclass{epl2} 

\title{Antisymmetric Tensor Fields in Codimension Two Brane-World}

\author{G. Alencar\inst{1} \and R. R. Landim\inst{2} \and M. O. Tahim\inst{1} \and R. N. Costa Filho\inst{2} \and K. C. Mendes\inst{3} }
\shortauthor{G. Alencar \etal}

\institute{                    
  \inst{1} Universidade
Estadual do Cear\'a, Faculdade de Educa\c c\~ao, Ci\^encias e Letras do Sert\~ao Central-R. Epit\'acio Pessoa, 2554, 63.900-000  Quixad\'{a}, Cear\'{a},  Brazil.\\
  \inst{2} Departamento de F\'{\i}sica, Universidade Federal do Cear\'{a} - 
Caixa Postal 6030, Campus do Pici, 60455-760, Fortaleza, Cear\'{a}, Brazil. \\
\inst{3} Universidade Estadual do Cear\'a, Departamento de F\'{\i}sica - Av.
Paranjana, 1700, 60740-000 Fortaleza,Cear\'{a}, Brazil}
\pacs{03.65.-w}{First pacs description}
\pacs{03.65.Ca}{Second pacs description}
\pacs{02.30.Ik}{Third pacs description}

\abstract{
In this work we consider the issue of localization of antisymmetric tensor fields of arbitrary rank in a D dimensional Space-time with a codimension two membrane. A string-like defect is used to simulate the membrane. The localization of massless and massive fields is found. The mass spectrum is infinitely degenerate for each mass level and this is solved by coupling the $q-$form to fermions.}

\begin{document}

\maketitle

\section{Introduction}

Higher dimensional theories are source of several ideas from the mathematical and physical point of view. Its diversity comes from the number of fields that naturally appears in their descriptions. If our universe is multidimensional then, in $D=4$, all of these fields should present some signal in the accelerators. This is enough to situate the study of localization of all sort of fields as an important aspect in physics of extra dimensions. These scenarios are those in which membranes or small compactified extra dimensions plays crucial role. They are naturally embedded in the formalism of superstring Theories.

As is well known, the quantum superstrings present in its spectrum a bunch of bosonic and fermionic tensorial fields: in fact, an infinite tower of them  \cite{Polchinski:1998rq,Polchinski:1998rr}. The understanding of the superstrings low energy limit is based on its massless excitations, and the low tension limit involves all the fields contained in its spectrum (in this case, all fields are massless)\cite{Becker:2007zj}. From this viewpoint, they are of great interest because they may have the status of fields describing particles other than the usual ones. As an example we can cite the space-time torsion \cite{Mukhopadhyaya:2004cc} and the axion field \cite{Arvanitaki:2009fg,Svrcek:2006yi} that have separated descriptions by the two-form. Other applications of these kind of fields are its relation with the AdS/CFT conjecture \cite{Germani:2004jf}.

The spectrum of the superstring is classified in various sectors. For example, the bosonic massless fields of the Ramond-Ramond sector in type IIA superstring is different from that in type IIB superstring. In type IIA we get a $1$-form and a $3$-form, while in type IIB we get a $0$-form, a $2$-form and a $4$-form. The existence of these fields is related to the existence of stable $D-$branes in type II theories. These $D-$branes are electrically or magnetically charged under these fields, i.e., the forms and its duals are important in the $D-$brane description.  In the context of $T-$duality in $D-$brane background, several kinds of $D-$branes are interrelated. Despite of these developments in the string viewpoint, there are studies about all characteristics of $p-$branes appearing in various SUGRAS.

From the mathematical side, antisymmetric tensors are natural objects living in the fiber bundle of differential manifolds. They play an important
role in the construction of the manifold's volume and therefore its orientation. The dimension of the manifold defines the sort of
possible antisymmetric tensors and this is used to define a specific space of forms \cite{Nakahara:2003nw}. Besides this, they are related to the linking number of higher dimensional knots \cite{Oda:1989tq}. Therefore, because of all these aspects, it is important to study higher rank tensor fields in membrane backgrounds. In this direction antisymmetric tensor fields has already been considered in models of extra dimension. Generally, the $q-$forms of highest order do not have physical relevance. This is due to the fact that when the number of dimensions increase, also increases the number of gauge freedom \cite{Hata:1980hn}. This can be used to cancel the dynamics of the field in the brane. The mass spectrum of the two and three form have been studied, for example, in Refs. \cite{Mukhopadhyaya:2004cc} and \cite{Mukhopadhyaya:2007jn} in a context of codimension one in five dimensions. After that, the coupling between the two and three forms with the dilaton was studied, in different contexts, in \cite{DeRisi:2007dn,Mukhopadhyaya:2009gp,Alencar:2010mi}. The study of this kind of coupling, inspired in string theory, is important in order to produce a process that, in principle, could be seen in LHC. This is a Drell-Yang process in which a pair quark-antiquark can give rise to a three(two)-form field, mediated by a dilaton.

In another direction soliton-like solutions are studied with increasing interest in physics, not only in Condensed Matter,
as in Particle Physics and Cosmology. In brane models, they are used as mechanisms of field localization, avoiding the
appearance of the troublesome infinities. Several kinds of defects in brane scenarios are considered in
the literature \cite{yves:a,yves:b,yves:c}. In these papers, the authors consider brane world models where the brane is supported
by a soliton solution to the baby Skyrme model or by topological defects available in some models. As an example, in a recent paper, a model is considered  for coupling fermions to brane and/or antibrane modelled by a kink antikink system \cite{yves:d}. The localization of fields in a framework that consider the brane as a kink has been studied for example in \cite{Bazeia:2008zx,Bazeia:2007nd,Bazeia:2004yw,Bazeia:2003aw}. In this context the present authors have studied the issue of localization and resonances of a three form field in \cite{Alencar:2010hs} and in a separate paper $q-$form fields are analysed in a codimension one $p-$brane \cite{preparation}. The aspects of the extra dimension  with codimension two has been recently studied. We cite for example \cite{Gogberashvili:2003xa,Gogberashvili:2003ys} for the gravitational field  where a new solution to the Einstein equations in 1+5 spacetime with an embedded 1+3 brane was presented. In \cite{Oda:2000zc} it was considered the localization of scalar and the vector gauge field in arbitrary dimension.
The subject of this article is to study aspects of $q-$form fields in a scenario of a $p-$brane with codimensions two. Specifically we will study the issue of localization using string-like solutions in a Randall-Sundrum-like model \cite{Oda:2000zc}.

This work is organized as follows. The second section is devoted to find the solution of Einstein
equation with source given by a string like defect. In the third section we analyse how the gauge freedom can be used to cancel the 
the angular component of an arbitrary $q-$form. In the fourth section we analyse the issue of localization of the zero modes of these fields . In the next section we solve the equation to find the mass spectrum and find that each mass level is infinitely degenerate. This is solved by coupling fermions to the fields. Finally, in the last section, we discuss the conclusions and perspectives.

\section{The Gravitational String-Like Solution}

As said in the introduction, topological defects has been widely considered
in Randall-Sundrum like models. Here we consider a generalization
of this models in which a vortex is considered as the source for the
gravitation field \cite{Oda:2000zj}.

The action regarded is given by
\begin{eqnarray}
S = \frac{1}{2 \kappa_D^2} \int d^D x
\sqrt{-g} \left(R - 2 \Lambda \right)
+ \int d^D x  \sqrt{-g} L_m,
\label{acao}
\end{eqnarray}
where we are using the same notation as that of \cite{Oda:2000zj}. In the above action $\kappa$ is related to the
$D$-dimensional gravitational constant through the relation $\kappa = 8\pi G_N$ and $L_m$ is the contribution of the vortex to the Lagrangian.

The equation of motion are given by
\begin{eqnarray}
R_{MN} - \frac{1}{2} g_{MN} R
= - \Lambda g_{MN}  + \kappa_D^2 T_{MN}.
\label{EM}
\end{eqnarray}
where $T_{MN}$ is the energy-momentum tensor. In order to solve the above equation we choose the following ansatz for the metrics
\begin{eqnarray}
ds^2 &=& g_{MN} dx^M dx^N  \nonumber \\
&=& e^{-A(r)} \eta_{\mu\nu} dx^\mu dx^\nu + dr^2
+ e^{-B(r)} d \theta^2,
\label{metrics}
\end{eqnarray}
where in the above equation $r$ and $\theta$ are respectively the radial and angular coordinates of the vortex. This metric is similar
to that used in \cite{Gogberashvili:2003xa,Gogberashvili:2003ys}.In all the above equations we have used $M,N...$ as $D-$dimensional indices and $\mu,\nu...$ as $p-$brane indices. Furthermore we adopt for $T_{MN}$ the
spherically symmetric ansatz
\begin{eqnarray}
T^\mu_\nu &=& \delta^\mu_\nu t_o(r),  \nonumber \\
T^r_r &=& -T^{\theta}_{\theta}=t_\theta (r).
\label{Energy}
\end{eqnarray}

As pointed in \cite{Oda:2000zj}, the above choice gives us the spontaneous symmetry breakdown \cite{Olasagasti:2000gx} and a global defect. With this ansatz at hand, the solution of Einstein equation is given by
$$
ds^{2}=e^{-cr}\eta_{\mu\nu}dx^{\mu}dx^{\nu}+dr^{2}+R_{0}^{2}e^{-c_{1}r}d\theta^{2}
$$
where
\begin{eqnarray}
c^2 &=& \frac{1}{p(p+1)}(-8 \Lambda + 8 \kappa_D^2 t_\theta) > 0, \nonumber \\
c_1 &=& c - \frac{8}{pc} \kappa_D^2 t_\theta.
\label{c}
\end{eqnarray}
and defining $P\left(r\right)=e^{-cr}$ and $Q\left(r\right)=R_{0}^{2}e^{-c_{1}r}$
it can be written as
$$
ds^{2}=P\eta_{\mu\nu}dx^{\mu}dx^{\nu}+dr^{2}+Qd\theta^{2}.
$$

The case without sources is obtained in a trivial way by setting $t_\theta = 0$ giving us a local defect \cite{Gherghetta:2000qi}. We are going to use the above metrics to analyse the localization of arbitrary forms in string-like defects.

\section{Antisymmetric Tensor Fields in Arbitrary Dimensions}

In this section we consider the action for a $q-$form $X_{M_{1}...M_{q}}$. The
zero, one, and two forms have already been considered in the literature \cite{Oda:2000zj}, and a discussion about the degrees of freedom using BRST formalism can be found in \cite{Hata:1980hn}. In order to analyse arbitrary
forms we must first understand the gauge freedoms involved. First, note that the number of degrees of freedom (d.o.f.) of an antisymmetric tensor field with $q$ indices is given by
$$
\left(\begin{array}{c}
D\\q
\end{array}\right)=\frac{D!}{q!\left(D-q\right)!}.$$

For the one form we have
$$
\delta X_{M}=\partial_{M}\phi$$
where $\phi$ is a scalar field. Therefore we have for the number of physical d.o.f. of the one form
\begin{eqnarray}
&&\left(\begin{array}{c}
D\\ \nonumber
1\end{array}\right)-\left(\begin{array}{c}
D\\
0\end{array}\right)=
\\
&=&\left[\left(\begin{array}{c}
D-1\\
1\end{array}\right)+\left(\begin{array}{c}
D-1\\
0\end{array}\right)\right]-\left(\begin{array}{c}
D-1\\
0\end{array}\right) \nonumber \\
&=&\left(\begin{array}{c}
D-1\\
1\end{array}\right). \label{one form}
\end{eqnarray}

In the above expression we have used the Stiefel relation
$$
\left(\begin{array}{c}
D\\
k\end{array}\right)=\left(\begin{array}{c}
D-1\\
k\end{array}\right)+\left(\begin{array}{c}
D-1\\
k-1\end{array}\right),$$
and the fact that
\begin{equation}
\left(\begin{array}{c}
D\\
0\end{array}\right)=\left(\begin{array}{c}
D-1\\
0\end{array}\right).
\end{equation}

For the two form we have
$$
\delta X_{M_{1}M_{2}}=\partial_{[M_{1}}X_{M_{2}]}$$
and therefore the gauge parameter is a one form with number of physical d.o.f.
given by Eq. (\ref{one form}). Using again the Stiefel relation we obtain
\begin{eqnarray}
&&\left(\begin{array}{c}
D\\
2\end{array}\right)-\left(\begin{array}{c}
D-1\\ \nonumber
1\end{array}\right)=\\ \nonumber
&=&
\left[\left(\begin{array}{c}
D-1\\
2\end{array}\right)+\left(\begin{array}{c}
D-1\\
1\end{array}\right)\right]-\left(\begin{array}{c}
D-1\\
1\end{array}\right) \nonumber \\
&=&\left(\begin{array}{c}
D-1\\
2\end{array}\right).
\end{eqnarray}

Therefore we find that the number of physical d.o.f. of the $q-$form field is given
by
\begin{equation}
\left(\begin{array}{c}
D-1\\
q\end{array}\right). \label{dof}
\end{equation}

Lets prove that this is valid for a $(q+1)-$form. In this case the gauge
parameter will be an $q-$form and therefore, using the Stiefel relation,
we have for the physical d.o.f. of the $(q+1)-$form
\begin{eqnarray*}
&&\left(\begin{array}{c}
D\\
q+1\end{array}\right)-\left(\begin{array}{c}
D-1\\
q\end{array}\right)=\\
&=&
\left[\left(\begin{array}{c}
D-1\\
q+1\end{array}\right)+\left(\begin{array}{c}
D-1\\
q\end{array}\right)\right]-\left(\begin{array}{c}
D-1\\
q\end{array}\right) \\
&=&\left(\begin{array}{c}
D-1\\
q+1\end{array}\right).
\end{eqnarray*}

The D-form has no dynamics. Using gauge symmetries, from the above result, we can see that the d.o.f. of the $(D-1)$ form can be made all null at the visible brane.
Therefore, in the next section, we must analyse only the cases $q=0,1,2,3...p$.

\section{Localization of $q-$Form Zero Modes}

The path way for dealing with an arbitrary form is the same, therefore we will left the number
of indices free, like above. The field strength is denoted by $Y_{M_{1}...M_{q+1}}=\partial_{\lbrack M_{1}}X_{...M_{q+1}]}$ and the action is give  by
\begin{eqnarray}
S=-\frac{1}{\left(q+1\right)!^2}\int d^{D}x\sqrt{-g}Y^{M_{1}...M_{q+1}}Y_{M_{1}...M_{q+1}}
\label{action x}
\end{eqnarray}
and for the equation of motion
$$
\frac{1}{\sqrt{-g}}\partial_{M_{1}}\left(\sqrt{-g}g^{M_{1}N_{1}}...g^{M_{q+1}N_{q+1}}Y_{N_{1}...N_{q+1}}\right)=0.$$
With the above background the equation of motion becomes
\begin{eqnarray}
&&P^{-1}\eta^{\mu_{1}\nu_{1}}g^{M_{2}N_{2}}...g^{M_{q+1}N_{q+1}}\partial_{\mu_{1}}Y_{\nu_{1}...N_{q+1}}+ \nonumber \\
&&P^{-\frac{p}{2}}Q^{-1/2}\partial_{r}\left(P^{\frac{p}{2}}Q^{1/2}g^{M_{2}N_{2}}...g^{M_{q+1}N_{q+1}}Y_{r...N_{q+1}}\right)
\nonumber+ \\
&&Q^{-1}g^{M_{2}N_{2}}...g^{M_{q+1}N_{q+1}}\partial_{\theta}Y_{\theta...N_{q+1}}=0
\end{eqnarray}

The number of d.o.f. of the component $X_{M_{1}...M_{q-1}\theta}$ is given by
\begin{equation}
\left(\begin{array}{c}
D-1\\
q\end{array}\right),
\end{equation}
and from our previous analysis, it can be gauged away or, explicitly $X_{M_{1}...M_{q-1}\theta}=0$. Decomposing now the $q-$form field as
\begin{eqnarray}
X_{\mu_{1}\mu_{2}...\mu_{q}} & = & \Sigma_{l,m}B^{l,m}_{\mu_{1}\mu_{2}...\mu_{q}}\left(x^{\mu}\right)\frac{\rho_{m}\left(r\right)}{\sqrt{R_0}}e^{il\theta}\nonumber \\
X_{r\mu_{2}...\mu_{q}} & = & \Sigma_{l}B^{l}_{r\mu_{2}...\mu_{q}}\left(x^{\mu}\right) \frac{\sigma\left(r\right)}{\sqrt{R_0}}e^{il\theta} \label{decomposition}
\end{eqnarray}
and the reason for not summing over $m$ in the last equation is that $\sigma$ do not depend on it, as will be seen later. We easily see that the $s$-wave$(l=0)$ with constant radial dependence
$\rho_{m}=\rho_{0}$ and $X_{r\mu_{2}...\mu_{q}}=constant$ is solution.
To find this we need to use $\partial^{\mu_{1}}B_{\mu_{1}\mu_{2}...\mu_{q}}=\partial^{\mu_{1}}H_{\mu_{1}\mu_{2}...\mu_{q+1}}=0$,
where $H_{\mu_{1}\mu_{2}...\mu_{q+1}}=\partial_{[\mu_{1}}B_{\mu_{2}...\mu_{q+1}]}$.

Now we must substitute this solution into our action in order to look for a localization of this zero mode.  With this framework the action reduces to
\begin{eqnarray*}
S & = & -\frac{1}{\left(q+1\right)!^2}\int d^{D}x\sqrt{-g}Y^{M_{1}...M_{q+1}}Y_{M_{1}...M_{q+1}}=\\
 & = & -\frac{2\pi}{\left(q+1\right)!^2 R_0}\int_{0}^{\infty}drP^{\frac{p}{2}-q-1}Q^{1/2}\times\\
&\times&\int d^{p}x\eta^{\mu_{1}\nu_{1}}...\eta^{\mu_{q+1}\nu_{q+1}}H_{\mu_{1}...\mu_{q+1}}H_{\nu_{1}...\nu_{q+1}.}
\end{eqnarray*}

Therefore, in order to have a localized field, the integral
$$
I=\frac{1}{R_0}\int_{0}^{\infty}drP^{\frac{p}{2}-q-1}Q^{1/2}$$
has to be finite. Using our definition we get
$$
I=\int dre^{-[(\frac{p}{2}-(q+1))c+\frac{1}{2}c_{1}]r}.$$

The condition for localization can be expressed as
$$
\frac{1}{\kappa_{D}^{2}}\Lambda<t_{\theta}<-\frac{p-1-2q}{2(q+1)\kappa_{D}^{2}}\Lambda,
$$
for $c>0$ and
$$
t_{\theta}>-\frac{p-1-2n}{2(q+1)\kappa_{D}^{2}},
$$
for $c<0$.

The above expression generalizes the one found in \cite{Oda:2000zc} and in particular for $q=0$ and $q=1$ we recover the previous results.
We also must be careful because for an even $p$ the $(p/2)-$form is self-dual, and the action considered above is not valid for these cases and must be considered separately. In the case of a local defect, the above condition reduces to $q<(p-1)/2$. Therefore, for $p=4$ for example, only the zero and one form are localized. In the next section we must consider the massive modes.

\section{The Massive Modes}
Now we must consider the possibility of localization of the massive modes. For this we must consider the local defect cited above.
\begin{eqnarray}
&&\left(\eta^{\mu\nu} \partial_\mu \partial_\nu
+ P^{\frac{2q+1-p}{2}} \partial_r P^{\frac{p-2q+1}{2}} \partial_r
+ \frac{1}{R_0^2} \partial_\theta^2 \right) B_{\mu_1...\mu_q}\nonumber\\ 
&&- P^{\frac{2q+1-p}{2}} \partial_r P^{\frac{p-2q+1}{2}} \partial_{\mu_1}
B_{r\mu_2...\mu_q} = 0, \\
&&\left(\eta^{\mu\nu} \partial_\mu \partial_\nu
+ \frac{1}{R_0^2} \partial_\theta^2 \right) B_{r\mu_2...\mu_q} = 0,  \\
&&\partial_r \left( P^{\frac{p-1+2q}{2}} \partial_\theta B_{r\mu_2...\mu_q}
\right) = 0.
\label{3 eq of motion}
\end{eqnarray}

Using the previous decomposition of the fields (\ref{decomposition}), the last equation give us
\begin{eqnarray}
\sigma(r) = \alpha P^{-\frac{p-1+2q}{2}},
\end{eqnarray}
where $\alpha$ is an integration constant. Therefore, as stated before, we see that $\sigma$ does not depends on $m$. Using this solution and (\ref{decomposition})
we find that the first two equations in (\ref{3 eq of motion}) reduces to
\begin{eqnarray}
&\left(\eta^{\mu\nu} \partial_\mu \partial_\nu
- m_m^2 \right) B^{l,m}_{\mu_{1}\mu_{2}...\mu_{q}}\left(x^{\mu}\right) = 0,& \nonumber \\
&\left(\eta^{\mu\nu} \partial_\mu \partial_\nu
- \frac{l^2}{R_0^2} \right) B^{l}_{r\mu_{2}...\mu_{q}}\left(x^{\mu}\right) = 0,&
\label{2 eq of motion}
\end{eqnarray}
provided $\rho_m$ satisfy the equation
\begin{eqnarray}
 \left( P^{\frac{2q+1-p}{2}} \partial_r P^{\frac{p-2q+1}{2}} \partial_r - \frac{l^2}{R_0^2} \right) \rho_m(r) = -m_m^2 \rho_m(r).
\end{eqnarray}
or
\begin{eqnarray}
P\partial_r^2 \rho_m(r) + P'(\frac{p-2q+1}{2} )\partial_r \rho_m(r)\nonumber\\ + (m_0^2 - \frac{l^2}{R_{0}^2} ) \rho_m(r) = 0.
\label{mass equation}
\end{eqnarray}

Using our KK decomposition (\ref{decomposition}), the above solution for $\sigma$ and 
\begin{eqnarray}
\int_{0}^{\infty} dr P^{\frac{p-(2q+1)}{2}}
\rho_n \rho_{n'} = \delta_{nn'}.
\label{orthonormality}
\end{eqnarray}
we find the following effective action
\begin{eqnarray}
S_X &=& \int d^p x \sum_{n,l=0}^{\infty} [-\frac{1}{(q+1)!^2}
\eta^{\mu_{1} \nu_{1}}...\eta^{\mu_{q+1}\nu_{q+1}} H_{\mu_{1}...\mu_{q+1}}^{(n,l)}
H_{\nu_{1}...\nu_{q+1}}^{(n,l)} \nonumber \\
&-& \frac{1}{2} m_n^2 \eta^{\mu_{1}\nu_{1}}...\eta^{\mu_{q}\nu_{q}} B_{\mu_{1}...\mu_{q}}^{(n,l)}
B_{\nu_{1}...\nu_{q}}^{(n,l)} ] \nonumber - \frac{\alpha^2}{c(p-(2q-1))} [P(\bar{r})^{-\frac{p-(2q-1)}{2}} - 1]\times \nonumber \\
&&\int d^p x \sum_{l=0}^{\infty} [-\frac{1}{(q)!^2}
\eta^{\mu_{1} \nu_{1}}...\eta^{\mu_{q}\nu_{q}} H_{\mu_{1}...\mu_{q}}^{(n,l)}
H_{\nu_{1}...\nu_{q}}^{(n,l)} - \nonumber \\
&& m_n^2 \eta^{\mu_{1}\nu_{1}}...\eta^{\mu_{q-1}\nu_{q-1}} B_{\mu_{r} \mu_{1}...\mu_{q-1}}^{(n,l)}
B_{\nu_{r} \nu_{1}...\nu_{q-1}}^{(n,l)} ],
\end{eqnarray}
where we must take the limit $\bar{r}\rightarrow \infty$ at the end.
 
We point here that the above orthonormality relation is possible only for $2q+1<p$, a condition that was also necessary for the massless case. We must also perform a field redefinition
$B^{l}_{r\mu_{2}...\mu_{q}} \rightarrow \alpha \sqrt{\frac{2}{c(p-(2q-1)}}\sqrt{P(\bar{r})^{-\frac{(p-(2q-1)}{2}} - 1} B^{l}_{r\mu_{2}...\mu_{q}}$ in order to absorb the divergence coming from 
the multiplicative factor in front of the second term in the effective action. Now we can solve mass equation (\ref{mass equation}) by a change of variables $M_n^2 = m_n^2
- \frac{l^2}{R_0^2}$, $z_n = \frac{2}{c} M_n P^{-\frac{1}{2}}$
and $h_n = P^{\frac{p-2q+1}{4}} \rho_n$, to obtain
\begin{eqnarray}
\left[ \frac{d^2}{dz_n^2} + \frac{1}{z_n}\frac{d}{dz_n}
+ \left\{1 - \frac{1}{z_n^2} \left(\frac{p-2q+1}{2} \right)^2 \right\}
\right] h_n = 0.
\label{Bessel}
\end{eqnarray}

This is a Bessel equation of order $(p-2q+1)/2$ and have already been solved, with the same boundary condition in \cite{Oda:2000zj}. To avoid repetition we must just use the results obtained 
by the author. The solution found is given by
\begin{eqnarray}
\rho_n (z_n) = \frac{1}{N_n} P^{-\frac{(p-2q+1)}{2}} \left[ J_{\frac{(p-2q+1)}{2}}
(z_n) + \alpha_n Y_{\frac{(p-2q+1)}{2}}(z_n) \right].
\label{Bessel solution}
\end{eqnarray}

In the limit $M_n << c$, the mass formula and the normalization constant $N_n$ are given by
\begin{eqnarray}
M_n = \frac{c}{2} (n + \frac{p-2q}{4} - \frac{1}{2}) \pi
e^{-\frac{1}{2} c \bar{r}},
\label{mass}
\end{eqnarray}  
\begin{eqnarray}
N_n = \sqrt{c} \frac{z_n(\bar{r})}{2M_n} 
J_{\frac{p-2q+1}{2}}(z_n(\bar{r})).
\label{normalization}
\end{eqnarray}

Note that for the case $q=0,1$ our expressions reproduces the one found in \cite{Oda:2000zj}. The important fact to be noted here is that in the limit $\bar{r} \rightarrow \infty$ we have $M_n=0$
and therefore $m_{l}^{2}=\frac{l^2}{R_{0}^{2}}$. This means that each mass level is degenerate in $n$ and we would have an infinite massless modes localized on the $p-$brane. This problem
can be solved when we study the coupling of these fields to fermions. As we will see, only for the case $n=0$ the interaction term is localized. We must consider for this only the cases $q>1$ for 
the cases with $q=0,1$ has already been studied \cite{Oda:2000zj} and because the coupling of the $q-$form with fermions is fundamentally different for $q>1$. The coupling we are going 
to consider is given by
\begin{eqnarray}
S_{\Psi\bar{\Psi}X} = -g_X\int d^D x \sqrt{-g} \bar{\Psi} \Gamma^{M_{1}...M_{q+1}}Y_{M_{1}...M_{q+1}} \Psi \delta(r),
\label{interaction}
\end{eqnarray}

Using now the fact that $\Gamma^\mu = P^{-\frac{1}{2}} \gamma^\mu$ and integrating in $\theta$ we obtain for the $q-$form interaction
\begin{eqnarray}
S_{\bar{\Psi}\Psi B} = - g_X \sqrt{R_0} \int d^p x \bar{\Psi} \gamma^{\mu_{1}...\mu_{q+1}} 
\sum_{n=0}^\infty H_{\mu_{1}...\mu_{q+1}}^{(n,0)}(x) f_n(0) \Psi.
\end{eqnarray}
 
From our solution $\rho_m$ and for $M_n<<c$ we obtain the approximation
\begin{eqnarray}
\rho_n(0) = \frac{1}{N_n} J_{\frac{p-2q+1}{2}}(\frac{2}{c} M_n) 
= \sqrt{c} P^{\frac{1}{4}}(\bar{r}),
\end{eqnarray}
and using the orthonormality condition we also obtain
\begin{eqnarray}
\rho_0 = \sqrt{\frac{c(p-2q+1)}{2}}.
\label{62}
\end{eqnarray}

Defining now $\tilde{g}_X = g_X \sqrt{\frac{c(p-2q+1)}{2} R_0}$ we obtain for the effective interaction of the $q-$form
\begin{eqnarray}
S_{\bar{\Psi}\Psi B} = - \tilde{g}_X \int d^p x \bar{\Psi} \gamma^{\mu_{1}...\mu_{q+1}} 
\left[H_{\mu_{1}...\mu_{q+1}}^{(0,0)}(x) + \sqrt{\frac{2}{p-2q+1}} P^{\frac{1}{4}}
(\bar{r}) \sum_{n=1}^\infty H_{\mu_{1}...\mu_{q+1}}^{(n,0)}(x) \right] \Psi.
\label{effective interaction}
\end{eqnarray}

As said before, only the $n=0$ mode is localized due the $p^{\frac{1}{4}}$ factor multiplying the other modes. Therefore we do not have the problem of infinitely degenerate massless modes.
\section{Conclusions}
In this paper we have studied the issue of localization of zero modes of arbitrary forms in a $D-$dimensional space with codimension two.
To reach this we have used a string-like topological defect in the codimension cited. We also performed an analysis of the physical degrees of freedom of a $q-$form in order to cancel the angular component of the fields. With this at hand we first studied the localization of the massless fields in the global string-like topological defect. In this framework all form fields are localized if a given condition is satisfied. The condition found here is very similar to and generalizes the one found by I. Oda \cite{Oda:2000zj} for the zero and one form. For the massive case we use a local string-like solution. A solution to the mass equation is found and the spectrum is infinitely degenerate for each mass level. In order to solve this problem, we coupled the antisymmetric fields to fermions. This was done in a minimal way, by coupling the fermions to the gauge invariant field strength of the respective form. With these we found that only the first level $n=0$ is localized, solving therefore the problem of degeneracy, getting a spectrum that have dependence only on $l$.     

\acknowledgments 
We would like to thank the Laboratorio de Simulacao de oleos pesados for its hospitality.
The authors would like to acknowledge the financial support provided by Funda\c c\~ao
Cearense de Apoio ao Desenvolvimento Cient\'\i fico e Tecnol\'ogico
(FUNCAP) and the Conselho Nacional de Desenvolvimento Cient\'\i fico e Tecnol\'ogico (CNPq).

This paper is dedicated to the memory of my wife  Isabel Mara (R. R. Landim)

\end{document}